\documentclass[aps,twocolumn,amsfonts,amsmath,amssymb,prb]{revtex4}

\usepackage{graphicx}
\usepackage{graphics}
\usepackage{amsmath}
\usepackage{amsfonts}
\usepackage{color}

\begin{document}
\title{Multi-scale modelling of supercapacitors: 
From molecular simulations to a transmission line model
}

\author{Clarisse Pean$^{1,2,3}$, Benjamin Rotenberg$^{1,3}$, Patrice Simon$^{2,3}$, Mathieu Salanne$^{1,3,4}$}
\affiliation{$^1$Sorbonne Universit\'es, UPMC Univ Paris 06, CNRS, Laboratoire PHENIX, F-75005, Paris, France}
\affiliation{$^2$CIRIMAT, UMR CNRS 5085, Universit\'e Paul Sabatier, F-31062 Toulouse, France}
\affiliation{$^3$R\'eseau sur le Stockage Electrochimique de l'Energie (RS2E), FR CNRS 3459, France}
\affiliation{$^4$Maison de la Simulation, USR 3441, CEA - CNRS - INRIA - Universit\'e Paris-Sud -Universit\'e de Versailles, F-91191 Gif-sur-Yvette, France}

\begin{abstract}
 We perform molecular dynamics simulations of a typical nanoporous-carbon based supercapacitors. The organic electrolyte consists in 1-ethyl-3-methyl--imidazolium and hexafluorophosphate ions dissolved in acetonitrile. We simulate systems at equilibrium, for various applied voltages. This allows us to determine the relevant thermodynamic (capacitance) and transport (in-pore resistivities) properties. These quantities are then injected in a transmission line model for testing its ability to predict the charging properties of the device. The results from this macroscopic model are in good agreement with non-equilibrium molecular dynamics simulations, which validates its use for interpreting electrochemical impedance experiments.
\end{abstract}

\maketitle

\section{Introduction} 

\indent 
Carbon electrodes-based supercapacitors are electrochemical energy storage devices characterized by good power performances. They can be charged/discharged on very short timescales (e.g. a few seconds), so that they are useful in many applications~\cite{Miller_2008_Science}. They also show exceptional cycle lifes, since they can sustain millions of cycles whereas batteries survive a few thousand at best. This is due to the molecular mechanism involved in the charge storage. It simply consists in the adsorption of the ions from the electrolyte on high surface area electrodes. Unlike Li-ion batteries, there is no limitation due to faradaic reactions and to the transport of electrons and ions inside a low conductivity material~\cite{Simon_2014_Science}. The volumetric capacitance, and hence the energy density of the devices, dramatically increases when using nanoporous carbon electrodes~\cite{Chmiola_2006_Science}. This effect can be optimized by matching the average pore size with the adsorbed ions dimension~\cite{Largeot_2008_JACS,Lin_2009_EA}, even if the pore size dispersity should be taken into account~\cite{kondrat_effect_2012}. However, the power density performances of such devices may be affected. In sub-nanometric pores, the ions become highly confined~\cite{Merlet_2013_NatComm}, which has raised the question whether one should expect detrimental transport limitations.  

In recent years, several simulation studies have aimed at understanding the dynamics of the ions at the molecular scale. Kondrat {\it et al.} have combined molecular dynamics with a phenomenological mean-field type model to study the charging of slit-pores with various wettabilities towards ionic liquids. They have shown that in the generic case (wettable pores), charging is a diffusive process~\cite{Kondrat_2013_JPCC,Kondrat_2014_NatMat}. Although the calculated diffusion coefficients for the ions are generally lower than in the bulk ionic liquids, they observed a voltage range in which they eventually became larger, due to the onset of collective modes~\cite{Kondrat_2014_NatMat}. More recently, He {\it et al.} have also oberved such effects in molecular dynamics simulations when the size of the slit-pores match the ionic dimensions~\cite{he2015c}. In the case of the adsorption of electrolytes in more complex nanoporous carbons with several types of pores, the enhancement of the diffusivity is not observed~\cite{Pean_2015_JACS,rajput2014a}. However the diffusion coefficient values remain reasonable since they are on average lower than the bulk liquid ones by one order of magnitude only, which explains the fast charging of these supercapacitors.

Injecting the information extracted from molecular-scale simulations into macroscopic models would be very valuable. For supercapacitors, the canonical model used to interpret electrochemical impedance spectra is the transmission line model introduced by de Levie~\cite{delevie1963a,delevie1989a}. It is an equivalent circuit which consists in an infinite succession of
 electrode slices composed of a resistance and a capacitor, that are connected together. In a recent work we have shown that it was possible to fit the variation of the total charge on the electrode with respect to time when charging a nanoporous carbon-based supercapacitor using non-equilibrium molecular dynamics simulations~\cite{Pean_2014_ACSNano}. The only fitting parameter was the in-pore resistivity.  Here we go further in that direction, by showing that the transmission line model can be fully parameterized without any fitting, simply by including physical quantities determined in equilibrium molecular dynamics simulations. The resulting model shows a very good agreement with non-equilibrium molecular dynamics, which validates its use for the sample sizes we consider. The simulated system consists in a generic electrolyte, i.e. 1-butyl-3-methylimidazolium and hexafluorophosphate ions [C$_4$mim$^+$][PF$_6^-$] dissolved in
acetonitrile at the concentration of 1.5~mol L$^{-1}$, and a realistic model of nanoporous carbide-derived carbons for the electrodes.

\section{Molecular dynamics simulations}
\begin{figure*}[ht!]
\includegraphics[width=\textwidth]{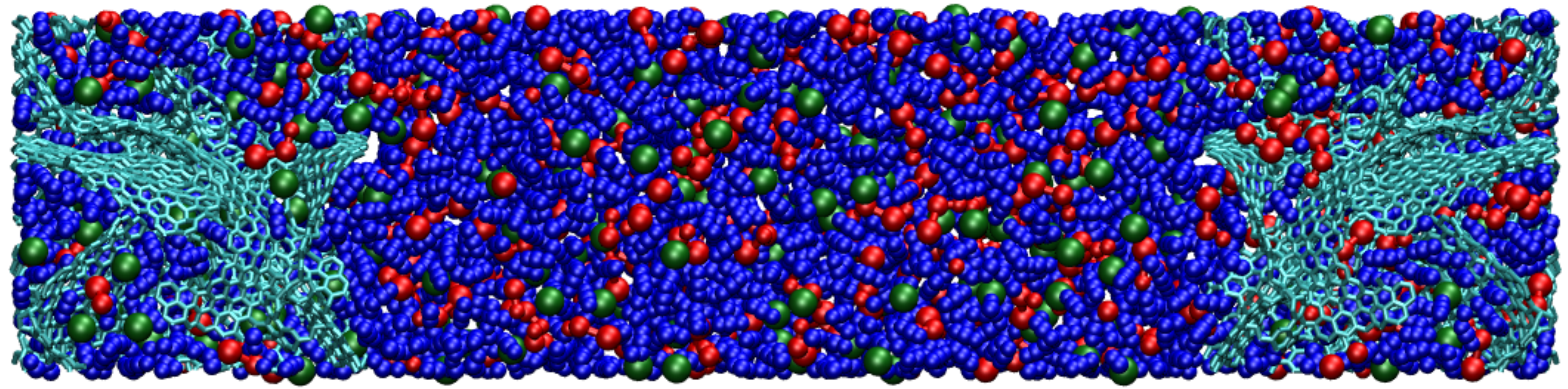}
\caption{\label{snapshot} Typical snapshot of the simulation cell. Blue: acetonitrile; red: C$_4$mim$^+$; green: PF$_6^-$ and turquoise: carbon electrodes.}
\end{figure*}

\begin{table}[h!]
\begin{center}
\begin{tabular}{|c|c|c|c|c|c|}
\hline
Simulations & $L_x=L_y$ (nm) & $L_z$ (nm) & N$_{\rm C}$ & N$_{\rm ions}$ & N$_{\rm ACN}$\\
\hline
Non-equilibrium & 4,37 & 19,44 & 3649 & 230 & 2146\\
\hline
Equilibrium & 4,37 & 32,46 & 3649 & 384 & 3584\\
\hline
\end{tabular}
\end{center}
\caption{\label{tableCDC}Lengths of the box and number of molecules for each
kind of simulation. The chosen carbide-derived carbon (CDC) is in both cases the CDC-1200 in contact with the ionic liquid [C$_4$mim$^+$][PF$_6^-$] dissolved in acetonitrile at a concentration of 1.5~mol L$^{-1}$. Two dimensional periodic boundary conditions are used, i.e. there is no periodicity in the $z$ direction. N$_{ \rm ions}$ corresponds to the number of ion pairs and N$_{\rm ACN}$ to the number of acetonitrile molecules.}
\end{table}

The simulation cells consists of [C$_4$mim$^+$][PF$_6^-$] dissolved in
acetonitrile (at the concentration of 1.5~mol L$^{-1}$, which is classically used
in experiments \cite{Tsai_2013_EC,Brandt_2014_JPS}), surrounded by two identical nanoporous
carbon electrodes placed symmetrically.  The position of the carbon atoms inside the electrodes, which are held fixed, were obtained by quenched molecular dynamics by Palmer \textit{et al.}~\cite{Palmer_2010_Carbon}. They have an average pore size of 0.9~nm, and the structural characteristics of the electrodes match well with the experimental data for a carbide-derived carbon synthesized at 1200~$^\circ$C. We will therefore label it CDC-1200 in the following.  The distances between the two electrodes along the $z$ direction are chosen in order to reproduce the experimental density of the bulk electrolyte. The lengths of the box are provided in Table \ref{tableCDC} together with the number of molecules, and a representative snapshot is shown in  Figure \ref{snapshot}.\\

Molecular dynamics simulations are conducted with a timestep of 2~fs. The simulations are performed in the NVE ensemble at room temperature (298~K). Following our previous works \cite{Merlet_2012_NatMat,Pean_2015_JECS}, we use the coarse-grained model of Roy and Maroncelli~\cite{Roy_2010_JPCBa, Roy_2010_JPCBb} for the ionic liquid [C$_4$mim$^+$][PF$_6^-$] and the one of Edwards \textit{et al.} \cite{Edwards_1984} for acetonitrile. Three interaction sites describe the cation and the acetonitrile molecules, while a single site describes the anion.  The Coulombic interactions are calculated through a two-dimensional Ewald summation~\cite{Reed_2007_JCP,Gingrich_2010} because two dimensional periodic boundary conditions are used (there is no periodicity in the $z$ direction). \\

Two series of simulations were performed. The first one involves
non-equilibrium simulations, which aim at simulating the charging process,
and proceeds via the following steps. The system is first equilibrated
for a few nanoseconds with a constant charge of 0~e on all carbon atoms and then
with a 0~V potential difference between the two electrodes. Then at $t=0$, this
potential difference $\Delta\Psi^0$ is suddenly set to 1~V  (on each electrode
we have $\Psi^+=\Delta\Psi^0/2$ and $\Psi^-=-\Delta\Psi^0/2$) and maintained
constant using a method developed by Reed \textit{et al.}~\cite{Reed_2007_JCP}
from the model of metallic carbon electrodes proposed by Siepmann and
Sprik~\cite{Siepmann_1995_JCP}. In this  approach, the charge distribution on each electrode is obtained by requiring that the potential experienced by each atom is equal to the preset electrode potential value at each molecular dynamics time step. It is
computationally expensive compared to constant charge simulations but
is compulsory for a realistic description of the charge repartition, which is rather broadly distributed,~\cite{Merlet_2012_NatMat,Limmer_2013_PRL} and of dynamic processes
. In particular we have shown that simulations in which the electrode charge distribution is held fixed yields unphysically high temperature increases during transient regimes~\cite{Merlet_2013_JPCL}. Simulations are continued long enough for the total
charge accumulated on the electrode to reach a plateau. Discharge
simulations are also performed, by suddenly setting the potential
difference to 0~V. In this work we perform a charge-discharge-charge cycle, for a total simulation time of 16~ns.

The second series deals with equilibrium simulations. In that case, the system is first equilibrated with a constant charge of 0~e, $\pm$0.005~e, or $\pm$0.01~e, which is equally distributed among electrode atoms. We previously showed that these charges are good starting values to initialize constant potential simulations at respectively 0, 1 or 2~V potential differences. Constant-potential production runs of 13~ns are then performed at these values.   

\section{Results and discussion}

\subsection{Electrosorption observation}

\begin{figure*}[ht!]
\includegraphics[width=1\textwidth]{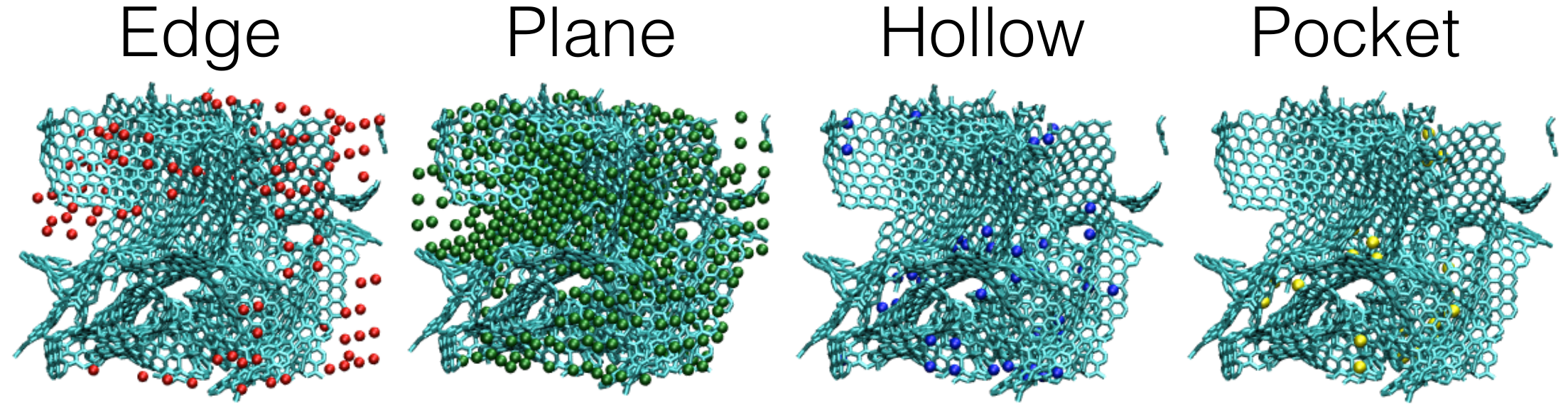}
\caption{\label{confinementsites} Repartition of the adsorption sites inside our modelled CDC electrode.}
\end{figure*}

It is well established that fluids have different transport properties when they
are confined in porous media compared to the
bulk~\cite{hummer2001a,Moore_2010_ASS,Farmahini_2014_JPCC,falk2015a,ori2015a}. In the case of electrolytes, the
diffusion coefficients of the ions are particularly
impacted~\cite{VanHaken_2014_JPCM, Cazade_2014}. In addition, when a liquid is
adsorbed inside a porous electrode with applied voltage, its
composition changes,sometimes leading to an unexpected enhancement of the diffusivity with respect to the bulk~\cite{Kondrat_2013_JPCC,Kondrat_2014_NatMat,he2015c}. In-pore resistivities, mean-square displacements, diffusion coefficients, as well as adsorption lifetimes are complementary quantities which can be extracted from computer simulations to characterize the singular transport in the pores of the material. 

In the case of complex porous carbons such as the CDC we study here, there are numerous confinement sites that cannot be represented by simple slit pores. In a previous work, we have identified four different types of sites, which are respectivelly labelled edge, plane, hollow or pocket, from the less confined to the more confined. The ranking of the sites is based on the degree of confinement, which is defined by the percentage of the solid angle around the ion which is occupied by the carbon atoms\cite{Merlet_2013_NatComm}). These sites are relatively homogeneously distributed inside the CDC-1200 electrode, as shown in Figure \ref{confinementsites}. 

\begin{figure}[ht!]
\includegraphics[width=\columnwidth]{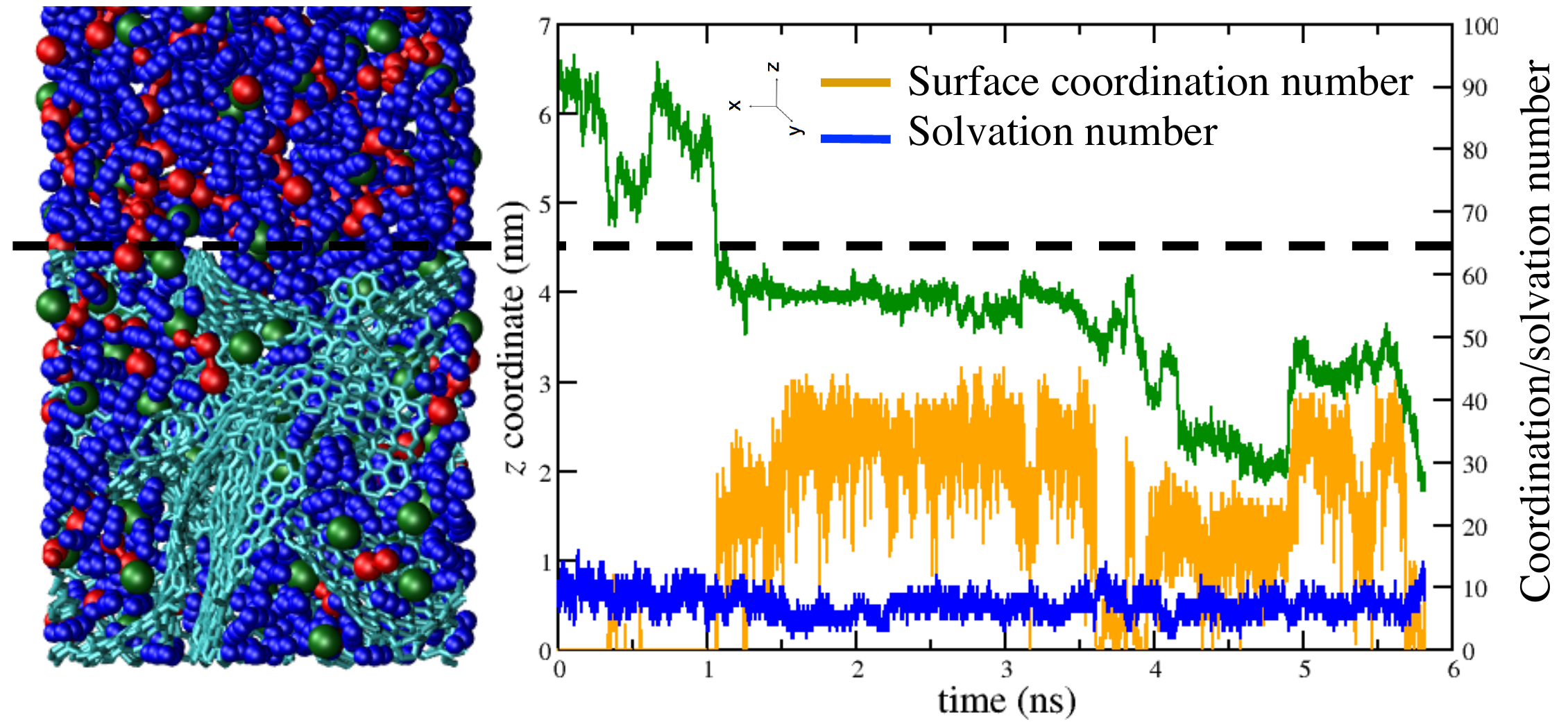}
\caption{\label{indiv}
Entrance of an anion inside the positive electrode : 
Evolution of the $z$ coordinate (green), of the solvation number (dark blue), 
of the surface coordination number (orange). The limit between the electrode and the \textit{bulk} is represented by the dashed line in black. The anion alternatively diffuses or becomes electrosorbed in pore sites of varying degree of confinement. When the number of coordination by the carbon increases, the number of solvation decreases.}
\end{figure}

In order to understand the charging mechanism in such a complex porous network,
it is instructive to examine the individual trajectories of the ions in the
simulation cell. An example is shown in Figure \ref{indiv} where we follow the
trajectory in the $z$ direction of an anion entering the positive electrode,
during a non-equilibrium charging simulation. During some time
intervals the ion is trapped - electrosorbed - at the electrode surface,
while during others it diffuses more freely. Electrosorption periods
are characterized by a fixed value of $z$ and of the ion coordination number by
the electrolyte molecule (solvation number, SN) and by the carbon atoms of the
electrode (surface coordination number, SCN), which are also shown on Figure
\ref{indiv}. When the ion is electrosorbed, the SCN fluctuates around a fixed
value corresponding to a hollow type  confinement site. When it diffuses, its
SCN decreases  while its SN increases. Thus, the carbon atoms counterbalance the
loss of solvent molecules inside the elctrode.  The typical residence times 
of electrosorbed ions at the surface, observed in Figure \ref{indiv}, are
around a nanosecond, which is consistent with the timescales
 in our previous work \cite{Pean_2015_JACS}. \\

\subsection{Transmission line model}

\begin{figure}[ht!]
\includegraphics[width=\columnwidth]{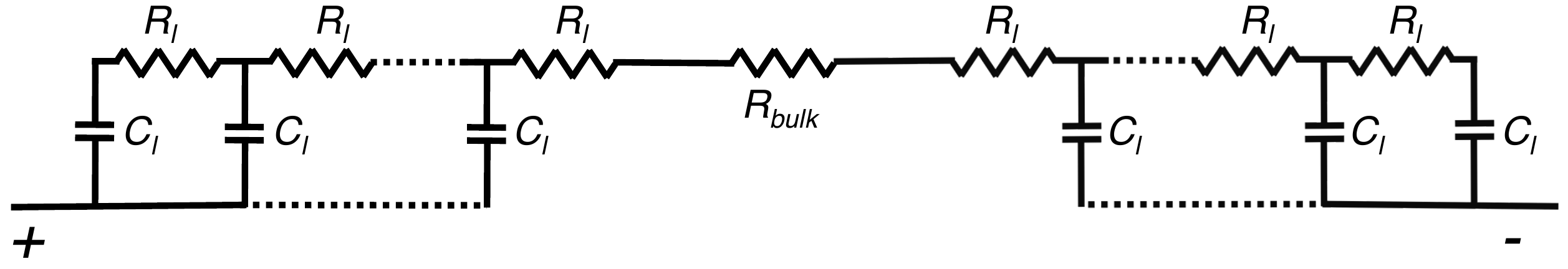}
\caption{\label{Circuit} Equivalent circuit in the transmission line model.
$R_{bulk}$ is the resistance of the electrolyte in the bulk region, while $R_l$
and $C_l$ are the resistance and the capacitance inside
the electrodes, respectively.}
\end{figure}

\indent When a potential difference is imposed to the supercapacitor, the
combination of all these individual diffusion/electrosorption
processes then leads to the global charging of the supercapacitor. The
interpretation of electrochemical impedance experiments generally relies on equivalent electric
circuits \cite{Zubieta_1997_TIA} and for nanoporous carbon electrodes, the most
used model is the transmission line model (TLM) \cite{delevie1963a,delevie1989a,Conway,
Taberna_2003_JES}, where the charge penetrates progressively inside the electrode. 
 We showed previously that this model was suitable to interpret our simulations
\cite{Pean_2014_ACSNano}. The TLM is based on an infinite succession of
electrode slices connected together, as shown in Figure \ref{Circuit}. Each slice is
described by a capacitance and a resistance. 

When modelling the simulation cell shown in Figure \ref{snapshot} with a TLM, the choice of the number of slices is delicate.
Indeed, the slices need to be as large as possible to account for a good
representation of the selected zones, meaning that the heterogeneities in the
charging process \cite{Pean_2014_ACSNano} due to the local porosity of the material and to the repartition of the confinement sites can be averaged out. Given the small size of the simulated
supercapacitors, we chose a two-slice TLM. In Figure \ref{Circuit}, $R_{bulk}$ is the resistance of the electrolyte in the layer of bulk liquid
between the two electrodes, $R_l$ is the resistance of the electrolyte adsorbed
in an electrode slice, and $C_l$ is the capacitance of each
slice. 

In this two-slice TLM, the total charge of the electrode is given by
the following expression \cite{Pean_2014_ACSNano}: 
\begin{equation}
Q(t)=Q_{max}\left[1-A_1 \exp\left(-\frac{t}{\tau_1}\right)-A_2 \exp\left(-\frac{t}{\tau_2}\right)\right]
\label{charge2tranches}
\end{equation}

where
\begin{eqnarray}
\label{eq:params1}
Q_{max} &=& \frac{c}{b}V_0 = C_l V_0
\nonumber \\ 
\tau_1 &=& \frac{2}{a + \sqrt{a^2-4b}} \nonumber \\ 
\tau_2 &=& \frac{2}{a - \sqrt{a^2-4b}}  \nonumber \\ 
A_1 &=& \frac{1}{2}\left[ 1 + \frac{ 2bd-ac }{ 2c \sqrt{a^2-4b}}\right]\nonumber \\ 
A_2 &=& \frac{1}{2}\left[ 1 - \frac{ 2bd-ac }{ 2c \sqrt{a^2-4b}}\right]
\; . \nonumber 
\end{eqnarray}
with
\begin{eqnarray}
\label{eq:params2}
a &=& \frac{2R_{bulk}+6R_l }
{ R_l(R_{bulk}+2R_l) C_l } \nonumber \\ 
b &=& \frac{ 2 }
{ R_l(R_{bulk}+2R_l) C_l^2 } \nonumber \\ 
c &=& \frac{2   }
{ R_l(R_{bulk}+2R_l) C_l } \nonumber \\ 
d &=& \frac{Ê1 }
{ R_{bulk}+2R_l } \; . \nonumber 
\end{eqnarray}

\ \\
\indent A very basic calculation allows to see how the components
of $Q(t)$ physically behave. In the approximation where $R_l$ is larger than
$R_{bulk}$, $A_1-1$ and $A_2-1$ scale as $R_l^{-1/2}$, while $\tau_1$ and $\tau_2$
scale as $R_l$.

The layers capacitance $C_l$ is a direct output from the simulations since we can measure the total charge accumulated on each region of the electrode at each timestep. $R_{\rm bulk}$ is computed from the electrical conductivity $\sigma =$ 5.3~S~m$^{-1}$. $R_l$ is the only parameter which is difficult to extract from the simulations. A first approach that we have used in a previous work~\cite{Pean_2014_ACSNano} is to determine it by fitting the plots $Q=f(t)$ obtained by non-equilibrium
simulations of the charging process using a least-square method. In our previous work, we have shown that the agreement
between the TLM and the raw data from the simulations is good, and so the two
slices model is sufficient for the supercapacitors modeled here.
  \\

In this work we describe another way to reach the unknown parameter $R_l$. From
equilibrium simulations conducted at a fixed potential value, the
transport of the anions, cations or acetonitrile molecules inside the electrodes
can be accurately characterized by computing their 
mean square displacements (MSD) and diffusion coefficients 
\cite{Pean_2015_JACS}.

\begin{figure}[ht!]
\includegraphics[width=\columnwidth]{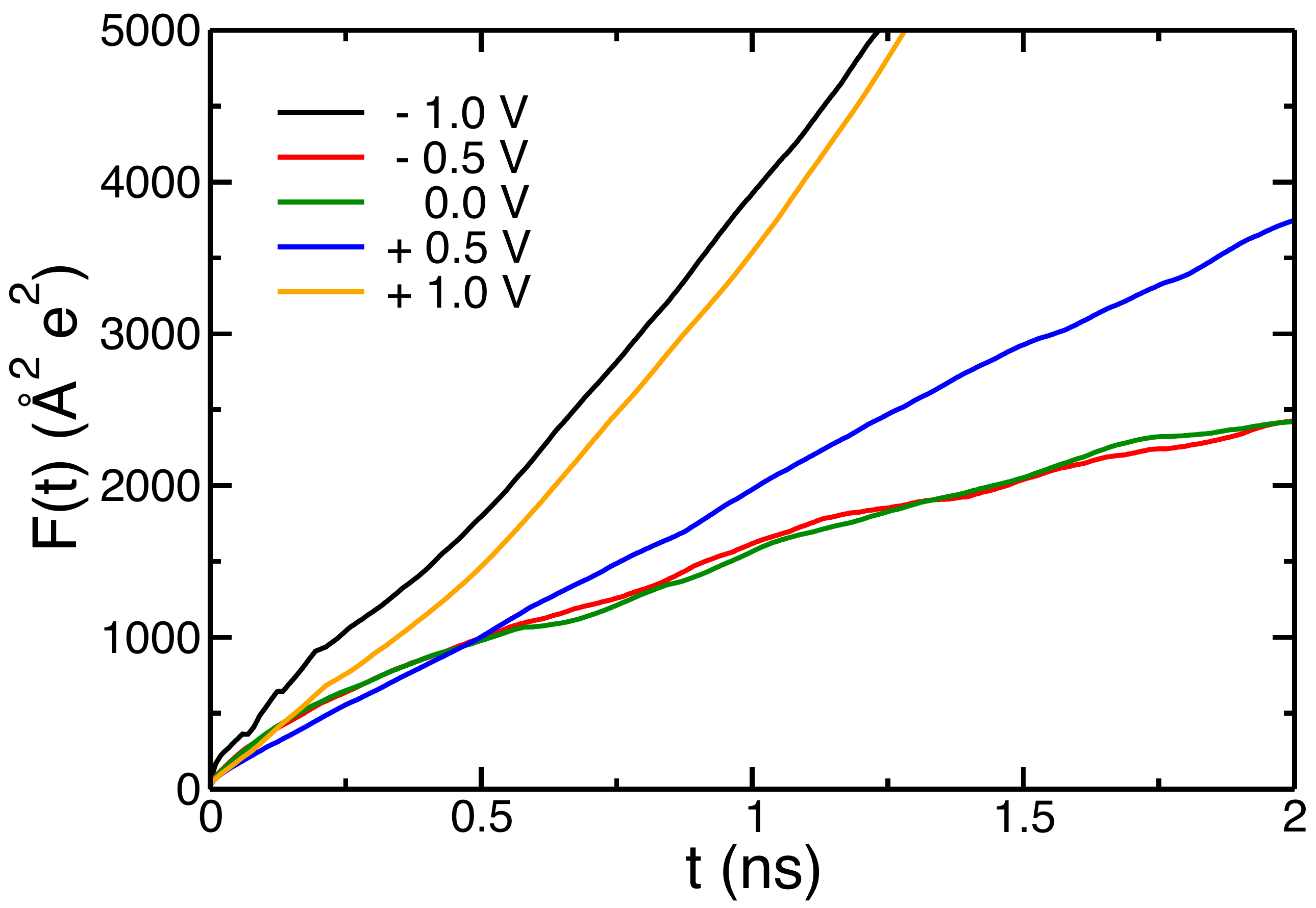}
\caption{\label{qmsd} Mean square displacements multiplied by anions and cations charge (Q-MSD), at different electrode potentials.}
\end{figure}

In particular, equilibrium molecular dynamics simulations allow to calculate the electrical conductivity of a bulk ionic liquid, by using the following Green-Kubo formula~\cite{hansen-livre}:
\begin{equation}
\label{eq:conductivitebulk}
K = \frac{\beta e^2}{V}\lim\limits_{t \to \infty} \frac{1}{6}\frac{\partial}{\partial t}\langle | \sum_i^N q_i \Delta \bm{r}_i(t) |^2 \rangle
\; ,
\end{equation}
where
$K$ is the conductivity, $\beta = 1/k_B T$
with $T$ the temperature and $k_B$ the Boltzmann constant, $V$ is the
volume of the system, $e$ the elementary charge, $\Delta
\bm{r}_i(t)$ is the displacement of ion $i$, bearing a formal charge $q_i$, during  time $t$ and $N$ is the number of ions in the simulation cell. Due to the sum over all the ions, long simulation times are required for gathering enough statistics. In the system we consider here, additional difficulties arise from the geometry of the simulation cell (two-dimensional periodicity) and the fact that the electrolyte is adsorbed in a porous network. We also need to account for the exchange of ions with the bulk electrolyte. We thus have to use the modified relationship
\begin{equation}
K' (\Psi^{\rm elec})= \frac{\beta e^2}{V_{\rm elec}}\lim_{N_t\rightarrow\infty}\frac{1}{4}\frac{\partial}{\partial t}F(t,\Psi^{\rm elec})
\label{eq:conductivite}
\end{equation}
\noindent where $F(t,\Psi^{\rm elec})$ is given by
\begin{equation}
F(t,\Psi^{\rm elec})\langle \mid \sum_{j}^{N_t} \sum_i^{N} q_i \delta {\bf r}^\perp_i(t_j-t_{j-1})S_{\rm D}^i(t_j)\mid^2\rangle_{\Psi^{\rm elec}}
\end{equation}
In this function, the two sums run respectively over the number of time intervals $N_t$ used to calculate the auto-correlation function and all the ions $N$ in the simulation cell.  $V_{\rm elec}$ is the volume occupied by the ions inside the electrode. $S_{\rm D}^i$ is a discontinuous presence function, which takes a value of 1 if an ion is present in the electrode at time $t_j$ and 0 otherwise. The displacements of ion $i$ are calculated along the $x$ and $y$ coordinates only, so that we note the corresponding quantity as $\delta {\bf r}^\perp_i$ (and the 6 in equation \ref{eq:conductivitebulk} is now replaced by a 4). The resulting in-pore conductivity depends on the electric potential which is applied to the electrodes $\Psi^{\rm elec}$.

\begin{figure}[ht!]
\includegraphics[width=\columnwidth]{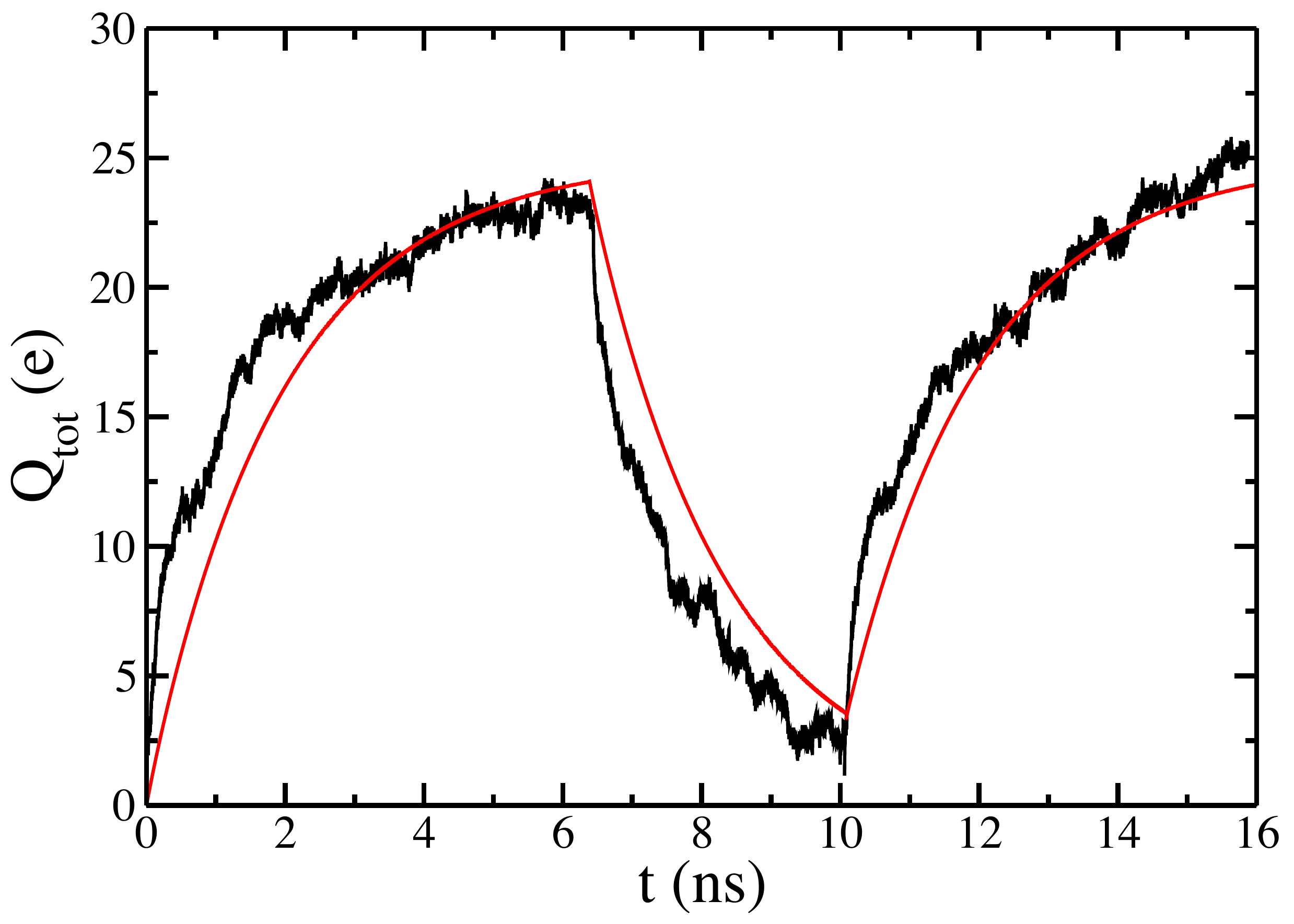}
\caption{\label{prediction_charges3} Comparison between the values of the total charge on the positive electrode during a charge-discharge cycle obtained with non-equilibrium simulations (black) and the predicted values of the total charge using the transmission line model using in-pore resistances $R_l$ obtained with equilibrium simulations (red).}
\end{figure}

The results obtained for the function $F(t,\Psi^{\rm elec})$ are reported for different electrode potentials in Figure
\ref{qmsd}. A linear r\'egime is reached for long times, from which we can determine $K' (\Psi^{\rm elec})$. The in-pore resistance defined in the transmission line model is then given by: 
\begin{equation}
R_l(\Psi^{\rm elec})= \frac{l}{A}\frac{1}{K' (\Psi^{\rm elec})}
\end{equation}
\noindent where $l$ is the length of the electrode slice (here it is the length of the electrode along the $z$ direction divided by 2) and $A$ its section. The dependence on the applied potential that we observe is not included in the TLM, so that we use the value of $R_l$ at 0~V in the following. When inserting this value into equation \ref{charge2tranches} ($R_l$ enters
in the constants $A_1$ et $A_2$ and mostly in the constants $\tau_1$ et $\tau_2$
, see equations \ref{eq:params1} and \ref{eq:params2}), we can calculate the $Q=f(t)$ plot for a charging process using equation \ref{charge2tranches} and the one for a discharging process using
\begin{equation}
Q(t)=Q_{max}\left[A_1 \exp\left(-\frac{t}{\tau_1}\right)+A_2 \exp\left(-\frac{t}{\tau_2}\right)\right]
\end{equation}

The  evolution of the total charge yielded by the  TLM is  compared in Figure \ref{prediction_charges3}  with the results of our non-equilibrium molecular dynamics simulations, for a charge-discharge-charge cycle. 
The agreement is remarkably good, 
which means that it is possible to predict in a reliable way the behaviour of supercapacitors in non-equilibrium conditions using equilibrium properties only. If we assume that the system is large enough to represent a macroscopic electrode, it validates the use of the TLM for interpreting experimental data and provides a microscopic insight on the parameters extracted from impedance spectroscopy.

\section{Conclusion}

The examination of individual trajectories of the electrolyte ions and
solvent molecules adsorbed inside a typical nanoporous carbide-derived carbon electrode during its charging shows that while they diffuse from pore to pore they experience many 
(electro-)sorption events at the electrode surface. When switching from the local pore scale to the macroscopic scale of an electrode, it is necessary to account for all the trajectories of all the adsorbed species. In terms of equivalent electrical circuit, such a behavior is well represented by the transmission line model introduced by de Levie.

In this work, we have shown that it is possible to fully parameterize a transmission line model from equilibrium molecular dynamics only. The relevant quantities are the bulk electrolyte conductivity and the capacitance, which are routinely evaluated in such simulations. The conductivity of the electrolyte {\it inside the electrodes} is more difficult to determine, and it requires to perform long simulations. The accuracy of the parameterized TLM is validated by comparing the evolution of the charge inside the electrode during a charge-discharge-charge cycle to the case of non-equilibrium molecular dynamics simulations. This study opens the door for a quantitative comparison between molecular-scale simulations and electrochemical impedance experiments, even if the results will have to be generalized to other systems. For example, many new electrolytes are currently designed in order to improve the energy density of supercapacitors~\cite{pohlmann2015b,schutter2015a}, and it will be necessary to determine their transport properties inside the pores for assessing their power performances. For the electrodes, many new materials are based on graphene~\cite{Gao_2014_Nanoenergy}, and are thus more ordered than the carbide-derived carbon we consider here. It will therefore be necessary to check the validity of the TLM under such conditions. Finally, in the case of pseudocapacitors~\cite{toupin2004a,brousse2015a,augustyn2013a}, the additional faradaic mechanisms may impact the charging dynamics at the molecular scale. A recent simulation study has shown that the electron transfer rates are enhanced by the frustration of the solvation structure close to MnO$_2$ surfaces~\cite{remsing2015a}. Determining such rates  would allow to derive macroscopic models for pseudocapacitors with deep molecular-scale insights.  

\section{Acknowledgments} 
We acknowledge the support from the European Research Council under the European Union's Seventh Framework Programme (FP/2007-2013) / ERC grant Agreement n.102539 (Advanced grant, Ionaces project). We are grateful for the computing resources on OCCIGEN (CINES, French National HPC) and CURIE (TGCC, French National HPC) obtained through the project x2015096728.

\section*{References}
\bibliography{biblio2}




\end{document}